\documentstyle[twoside,fleqn,espcrc2hep]{article}

\def\t{\textstyle}

\def\sst{\scriptscriptstyle}

\newcommand{\AmS}{{\protect\the\textfont2
  A\kern-.1667em\lower.5ex\hbox{M}\kern-.125emS}}

\title{
\vspace*{-2.8cm}
\begin{flushright}
{\normalsize LNF-00/20(P)\\[-05pt]
DESY-00-130}
\end{flushright}
\vspace{1.3cm}
$1/N_c$ and $\varepsilon'/\varepsilon$ \thanks{Invited talk  
presented by T.H. at QCD 00, July 6-13th 2000, Montpellier, France}}

\author{T.\ Hambye\address{INFN - Laboratori Nazionali di Frascati, 
        P.O. Box 13, I-00044 Frascati, Italy}
        and 
        P.H.\ Soldan\address{II.\ Institut f\"ur Theoretische Physik,
        Universit\"at Hamburg, \\ Luruper Chaussee 149, D-22761 Hamburg,
        Germany}
}

\begin{document}

\begin{abstract}
We present a recent analysis of $\varepsilon'/\varepsilon$ in the 
$1/N_c$ expansion. We show that the $1/N_c$ corrections to the matrix 
element of $Q_6$ are large and positive, indicating a $\Delta I=1/2$ 
enhancement similar to the one of $Q_1$ and $Q_2$ which dominate the 
CP conserving amplitude. This enhances the CP ratio and can bring the 
standard model prediction close to the measured value for central 
values of the parameters. Several comments on the theoretical status of 
$\varepsilon'/\varepsilon$ and the errors in its calculation are given.
\end{abstract}

\maketitle

\section{INTRODUCTION}

Direct CP violation in $K\rightarrow\pi\pi$ decays was recently
observed by the KTeV and NA48 collaborations \cite{ktev,fanti}. The 
present world average \cite{fanti} for the parameter $\varepsilon'/
\varepsilon$ is $\mbox{Re}\,\varepsilon'/\varepsilon\,=\,(19.3 \pm 
2.4)\cdot 10^{-4}$. In the standard model CP violation originates 
in the CKM phase, and direct CP violation is governed by loop diagrams 
of the penguin type. The main source of uncertainty in the calculation 
of $\varepsilon'/\varepsilon$ is the QCD non-perturbative contribution 
related to the hadronic nature of the $K\rightarrow\pi\pi$ decay. Using 
the $\Delta S=1$ effective hamiltonian,
\begin{equation}
{\cal H}_{ef\hspace{-0.5mm}f}^{\sst\Delta S=1}=\frac{G_F}{\sqrt{2}}
\;\lambda_u\sum_{i=1}^8 c_i(\mu)\,Q_i(\mu)\hspace{5mm}(\mu < m_c)\,,
\label{ham}
\end{equation}
the non-perturbative contribution, contained in the hadronic matrix 
elements of the four-quark operators $Q_i$, can be separated from the 
perturbative Wilson coefficients $c_i(\mu)=z_i(\mu)+\tau y_i(\mu)$ 
(with $\tau=-\lambda_t/\lambda_u$ and $\lambda_q=V_{qs}^*\,V_{qd}^{}$). 
Introducing $\langle Q_i\rangle_I\equiv\langle (\pi\pi)_I|Q_i|K\rangle$, 
the CP ratio can be written as
\begin{eqnarray}
\frac{\varepsilon'}{\varepsilon}&=&\frac{G_F}{2}
\frac{\omega\,\mbox{Im}\lambda_t}{|\varepsilon|\,\mbox{Re}A_0}
\Big[\,\Big|\sum_i\,y_i\,\langle Q_i\rangle_0\Big|\,
\Big(1-\Omega_{\mbox{\tiny IB}}\Big)\nonumber\\
&&-\,\frac{1}{\omega}
\Big|\sum_i\,y_i\,\langle Q_i\rangle_2\Big|\,\Big]\,.
\label{epspsm}
\end{eqnarray}
$\omega=$Re$A_0/$Re$A_2=22.2$ is the ratio of the CP conserving $K
\rightarrow\pi\pi$ isospin amplitudes; $\Omega_{\mbox{\tiny IB}}$ 
parameterizes isospin breaking corrections \cite{ecker}. 
$\varepsilon'/\varepsilon$ is dominated by $\langle Q_6\rangle_0$ and 
$\langle Q_8\rangle_2$ which cannot be fixed from the CP conserving data 
\cite{BJM,bosch}. Beside the theoretical uncertainties coming from the 
calculation of the $\langle Q_i\rangle_I$ and of $\Omega_{\mbox{\tiny IB}}$, 
the analysis of the CP ratio suffers from the uncertainties on the values of 
various input parameters, in particular of the CKM phase in Im$\lambda_t$, 
of $\Lambda_{\mbox{\tiny QCD}}\equiv\Lambda^{(4)}_{\overline{\mbox{\tiny 
MS}}}$, and of the strange quark mass. 

\section{\boldmath ON THE COUNTING IN $1/N_c$ AND THE USE 
OF THE LARGE-$N_c$ VALUES FOR THE MATRIX ELEMENTS \unboldmath}

To calculate the hadronic matrix elements we start from the effective
chiral lagrangian for pseudoscalar mesons which involves an expansion 
in momenta where terms up to ${\cal O}(p^4)$ are included \cite{GaL}. 
Keeping only terms of ${\cal O}(p^4)$ which contribute, at the order 
we calculate, to the $K \rightarrow \pi \pi$ amplitudes, for the 
lagrangian we obtain:
\begin{eqnarray}
\lefteqn{{\cal L}_{ef\hspace{-0.5mm}f}\,=} &&\nonumber\\[1mm]
&&
\hspace*{-3mm}\t \frac{f^2}{4}\big(\langle D_\mu U^\dagger D^{\mu}U\rangle
+\frac{\alpha}{4N_c}\langle \ln U^\dagger -\ln U\rangle^2 \nonumber\\[1mm] 
&&
\hspace*{-3mm}\t +\langle\chi U^\dagger+U\chi^\dagger\rangle\big) +L_8\langle 
\chi^\dagger U\chi^\dagger U+\chi U^\dagger\chi U^\dagger \rangle  
\nonumber\\[1mm]
&&
\hspace*{-3mm}+L_5\langle D_\mu U^\dagger D^\mu U(\chi^\dagger U
+U^\dagger\chi)\rangle\,,\label{lag}
\end{eqnarray}
with $\langle A\rangle$ denoting the trace of $A$, $\alpha=m_\eta^2
+m_{\eta'}^2-2m_K^2$, $\chi=r{\cal M}$, and ${\cal M}=\mbox{diag}
(m_u,m_d,m_s)$. $f$ and $r$ are parameters related to the pion decay 
constant $F_\pi$ and to the quark condensate, with $r=-2\langle\bar{q}q
\rangle/f^2$. The complex matrix $U$ is a non-linear representation of 
the pseudoscalar meson nonet. The conventions and definitions we use 
are the same as those in~\cite{HKPS,HKPSB,hks}. 
The method we use is the $1/N_c$ expansion \cite{BBG2}. In this 
approach, we expand the matrix elements in powers of the momenta and 
of $1/N_c$. For the $1/N_c$ corrections we calculated chiral loops as 
described in refs.~\cite{HKPSB,hks}. Especially important to this analysis 
are the non-factorizable corrections, which are UV divergent and must 
be matched to the short-distance part. They are regularized by a 
finite cutoff $\Lambda_c$ which is identified with the short-distance 
renormalization scale. The definition of the momenta in the loop 
diagrams, which are not momentum translation invariant, is discussed 
in detail in ref.~\cite{HKPSB}. Other recent work on matrix elements 
in the $1/N_c$ approach can be found in refs.~\cite{BP,BP2,kpr}. 

For the Wilson coefficients we use the leading logarithmic and the 
next-to-leading logarithmic values \cite{BJM}. The absence of any 
reference to the renormalization scheme in the low-energy calculation, 
at this stage, prevents a complete matching at the next-to-leading 
order \cite{ab98}. Nevertheless, a comparison of the numerical results 
obtained from the LO and NLO coefficients is useful as regards estimating 
the uncertainties and testing the validity of perturbation theory.

As it is well known the large-$N_c$ approximation fails completely in 
explaining the $\Delta I = 1/2$ rule. By taking the large-$N_c$ values 
for the two dominant operators $Q_1$ and $Q_2$ one obtains a $\Delta I=1/2$ 
CP-conserving amplitude which underestimates the data by roughly a factor 
of three. However, as pointed out in ref.~\cite{BBG2}, $Q_{1,2}$ show an
important specificity; they are expected to be largely affected by 
non-chirally suppressed corrections beyond the large-$N_c$ limit. 
In the counting in $p^2$ and $1/N_c$, this property is attributed to
the fact that the tree level mesonic representation of $Q_{1,2}$ from 
the leading ${\cal O}(p^2)$ chiral lagrangian introduces ${\cal O}(p^2)$ 
terms with two derivatives. As a result the one-loop contributions over these 
terms, which are ${\cal O}(p^2/N_c)$, are quadratically divergent, i.e.\ 
possibly very large because these corrections are not protected by any 
symmetry like the chiral one. Calculating the terms of ${\cal O}(p^2/N_c)$ 
one obtains a large enhancement of the $\Delta I = 1/2$ amplitude \cite{BBG2} 
in the range required to reproduce the experiment \cite{hks}. This illustrates 
how important are the non-chirally suppressed $1/N_c$ corrections for an 
understanding of the $K \rightarrow \pi \pi$ amplitudes.

In this context, for $\varepsilon'/\varepsilon$, it is very important to
investigate the counting in $p^2$ and $1/N_c$ for the dominant operators 
$Q_{6,8}$ and to compare it with the one for $Q_{1,2}$. For $Q_8$ the mesonic 
representation from the ${\cal O}(p^2)$ lagrangian is ${\cal O}(p^0)$ and does
not have any derivative (because $Q_8$ is a density-density operator whereas
$Q_{1,2}$ are of the current-current type). As a result we do not expect any 
quadratic term but only chirally suppressed logarithmic or finite terms. 
The large-$N_c$ limit, $B_8^{(3/2)}=1$, is therefore expected to be a much 
better approximation than for $Q_{1,2}$. Different is the case of $Q_6$ 
due to the fact that there is no tree level contribution from the leading 
${\cal O}(p^2)$ lagrangian, and the large-$N_c$ value $B_6^{(1/2)}=1$ 
refers to the tree level contribution from the ${\cal O}(p^4)$ 
lagrangian which, as for $Q_{1,2}$, carries two derivatives [i.e.\ is 
${\cal O}(p^2)$]. Therefore, as for $Q_{1,2}$, we expect the large-$N_c$
value of $B_6^{(1/2)}$ to be affected by large ${\cal O}(p^2/N_c)$ corrections 
[$\sim {\cal O}(100\,\%)$] resulting from the loops over the ${\cal O}(p^2)$ 
tree operator. This shows clearly that, in the same way as for the $\Delta 
I = 1/2$ rule, no clear statement can be done on the expected size of 
$\varepsilon'/\varepsilon$ without calculating these $1/N_c$ non-factorizable
corrections.

\section{ANALYSIS OF \boldmath $\varepsilon'/\varepsilon$\unboldmath}

Analytical formulas for all matrix elements, at next-to-leading order 
in the twofold expansion in powers of momenta and of $1/N_c$, are given in 
refs.~\cite{HKPSB,hks}. In the pseudoscalar approximation, the matching has 
to be done below 1\,GeV. Varying $\Lambda_c$ between 600 and 900\,MeV, 
the bag factors $B_1^{(1/2)}$ and $B_2^{(1/2)}$ take the values 
$8.2-14.2$ and $2.9-4.6$; quadratic terms in $\langle Q_1\rangle_0$ 
and $\langle Q_2\rangle_0$ produce a large enhancement which brings 
the $\Delta I=1/2$ amplitude in agreement with the data \cite{hks}. 
Corrections beyond the chiral limit were found to be small.

For $\langle Q_6\rangle_0$ and $\langle Q_8\rangle_2$ the leading
non-factorizable loop corrections, which are of ${\cal O}(p^0/N_c)$, are 
only logarithmically divergent \cite{HKPSB}. Including terms of ${\cal O}(p^0)$,
${\cal O}(p^2)$, and ${\cal O}(p^0/N_c)$, $B_6^{(1/2)}$ and $B_8^{(3/2)}$ take 
the values $1.10-0.72$ and $0.64-0.42$. As a result, as with large-$N_c$ 
values, $\varepsilon'/\varepsilon$ is obtained generally much smaller than the 
data, except for input parameters taken close to the extreme of the ranges we 
considered. However, as stated above, since the leading ${\cal O}(p^0)$ 
contribution vanishes for $Q_6$, corrections from higher order terms beyond 
the ${\cal O}(p^2)$ and ${\cal O}(p^0/N_c)$ are expected to be large. 
In ref.~\cite{HKPS} we investigated the ${\cal O}(p^2/N_c)$ contribution, 
i.e., the $1/N_c$ correction at the next order in the chiral expansion, 
because it brings about, for the first time, quadratic corrections on 
the cutoff. From counting arguments and more generally from the fact that 
the chiral limit is assumed to be reliable, the quadratic terms (which 
are not chirally suppressed) are expected to be dominant. It is still 
desirable to check that explicitly by calculating the corrections 
beyond the chiral limit, from logarithms and finite terms, as done 
for $Q_1$ and $Q_2$. Numerically, we observe a large positive correction 
from the quadratic term in $\langle Q_6\rangle_0$. This point was 
already emphasized in ref.~\cite{orsay}. The slope of the correction 
is qualitatively consistent and welcome since it compensates for the 
logarithmic decrease at ${\cal O}(p^0/N_c)$. Varying $\Lambda_c$ 
between 600 and 900\,MeV, the $B_6^{(1/2)}$ factor takes the values 
$1.50-1.62$. $Q_6$ is a $\Delta I=1/2$ operator, and the enhancement 
of $\langle Q_6\rangle_0$ indicates that at the level of the $1/N_c$ 
corrections the dynamics of the $\Delta I=1/2$ rule applies to $Q_6$ 
as to $Q_1$ and $Q_2$. The size of the enhancement for $B_6^{(1/2)}$  
appears however to be smaller for $Q_6$ than for $Q_{1,2}$ due to a 
smaller coefficient of the quadratic term. This coefficient is 
nevertheless large enough to increase $\varepsilon'/\varepsilon$ 
by almost a factor of two.   

Using the quoted values for $B_6^{(1/2)}$ together with the full leading 
plus next-to-leading order $B$ factors for the remaining operators 
\cite{HKPS} the results we obtain for $\varepsilon'/\varepsilon$ are 
given in tab.~\ref{tab1} for the three sets of Wilson coefficients LO, 
NDR, and HV and for $\Lambda_c$ between 600 and $900\,\mbox{MeV}$.
The numbers are close to the measured value for central values of the 
parameters (upper set). They are obtained by assuming zero phases from 
final state interactions. This approximation is very close to the results 
we would get if we used the small imaginary part obtained at the one-loop 
level \cite{HKPS}. 
\noindent
\begin{table}[t]
\caption{Numerical values for $\varepsilon'/ \varepsilon$ (in units of
$10^{-4}$) as explained in the text.
\label{tab1}}
\begin{eqnarray*}
\begin{array}{ccc}\hline
\rule{0cm}{5mm}
\mbox{LO}
&\,\,\,14.8 \,\,\leq\,\,\varepsilon'/\varepsilon\,\,\leq\,\,19.4 \,\,\,&
\\[0.2mm]
\mbox{\,\,NDR\,\,}
&\,\,\,12.5 \,\,\leq\,\,\varepsilon'/\varepsilon\,\,\leq\,\,18.3 \,\,\,& 
\mbox{central}
\\[0.2mm]
\mbox{HV}
& \,\,\,7.0\,\,\,\,\leq\,\,\varepsilon'/\varepsilon\,\,\leq\,\,14.9 \,\,\,&
\\[2.4mm]
\hline
\rule{0cm}{5mm}
\mbox{LO}
& \,\,\,6.1 \,\,\leq\,\,\varepsilon'/\varepsilon\,\,\leq\,\,48.5 \,\,\,&
\\[0.2mm]
\mbox{\,\,NDR\,\,}
& \,\,\,5.2 \,\,\leq\,\,\varepsilon'/\varepsilon\,\,\leq\,\,49.8 \,\,\,& 
\mbox{scanning}
\\[0.2mm]
\mbox{HV}
& \,\,\,2.2\,\,\leq\,\,\varepsilon'/\varepsilon\,\,\leq\,\,38.5 \,\,\, 
\\[2.4mm]
\hline    
\end{array}
\end{eqnarray*}
\end{table}

Performing a scanning of the parameters [$125\,\mbox{MeV}\leq m_s(1\,
\mbox{GeV})\leq 175$ $\mbox{MeV}$, $0.15\leq\Omega_{\mbox{\tiny IB}}\leq 0.35$, 
$1.04\cdot 10^{-4}\leq\mbox{Im}\lambda_t\leq 1.63\cdot 10^{-4}$, and $245
\,\mbox{MeV}\leq\Lambda_{\mbox{\tiny QCD}}\leq 405\,\mbox{MeV}$] we obtain 
the numbers in lower set of tab.~\ref{tab1}. They can be compared with the 
results of refs.~\cite{bosch,BP2,buras2,BEF,CM2}. The values of $B_8^{(3/2)}$
can also be compared with refs.~\cite{kpr,don}. Other recent calculations
are reported in refs.~\cite{bel,eap,narison}. The wide ranges reported in 
the table can be traced back, to a large extent, to the large ranges of the input 
parameters. This can be seen by comparing them with the relatively narrow 
ranges obtained for central values of the parameters. The parameters, to a 
large extent, act multiplicatively, and the large range for $\varepsilon'/
\varepsilon$ is due to the fact that the central value(s) for the ratio are 
enhanced roughly by a factor of two compared to the results obtained with 
$B$ factors for $Q_6$ and $Q_8$ close to the VSA. More accurate information 
on the parameters, from theory and experiment, will restrict the values for 
$\varepsilon'/\varepsilon$. 

To estimate the uncertainties due to higher order final state interactions 
we also calculated $\varepsilon'/\varepsilon$ using the real part of the 
matrix elements and the phenomenological values of the phases \cite{phases}, 
$\delta_0=(34.2\pm 2.2)^\circ$ and $\delta_2=(-6.9 \pm 0.2)^\circ$, i.e., 
we replaced $|\sum_i y_i\langle Q_i\rangle_I|$ in Eq.~(\ref{epspsm}) by 
$\sum_i y_i\mbox{Re}\langle Q_i\rangle_I/\cos\delta_I$. The corresponding
results are given in tab.~\ref{tab2}. They are enhanced by $\sim 25\,\%$ 
compared to the numbers in tab.~\ref{tab1}. To reduce the FSI uncertainties 
in the $1/N_c$ approach it would be interesting to investigate the two-loop 
imaginary part. By doing so we expect to get phases very close to the ones 
of ref.~\cite{GM} which have been obtained in Chiral Perturbation Theory at 
the same order and reproduce relatively well the data. In this sense
we expect to get results close to the ones of tab.~\ref{tab2}. A comparison 
of tabs.~\ref{tab1} and~\ref{tab2} will be however still useful to estimate 
higher order corrections (e.g.\ for the real part). In our analysis part of 
the uncertainty from higher order corrections is also included in the range 
due to the (moderate) residual dependence on the matching scale. In order to 
reduce the scheme dependence in the result, appropriate subtractions would 
be necessary \cite{BP,BP2,BB}. Finally, it is reasonable to assume that the 
effect of the pseudoscalar mesons is the most important one. Nevertheless, 
the incorporation of vector mesons and higher resonances would be desirable 
in order to improve the treatment of the intermediate region around the rho 
mass and to show explicitly that the large enhancement we find at low energy 
at the level of the pseudoscalars remains up to the scale $\sim m_c$, where 
the matching with the short-distance part can be done more safely. 
\noindent
\begin{table}[t]
\caption{Same as in Tab.~\ref{tab1}, but now with the phenomenological 
values for the phases.
\label{tab2}}
\begin{eqnarray*}
\begin{array}{ccc}\hline
\rule{0cm}{5mm}
\mbox{LO}
&\,\,\,19.5 \,\,\leq\,\,\varepsilon'/\varepsilon\,\,\leq\,\,24.7 \,\,\,&
\\[0.2mm]
\mbox{\,\,NDR\,\,}
&\,\,\,16.1 \,\,\leq\,\,\varepsilon'/\varepsilon\,\,\leq\,\,23.4 \,\,\,&
\mbox{central}
\\[0.2mm]
\mbox{HV}
& \,\,\,9.3\,\,\,\,\leq\,\,\varepsilon'/\varepsilon\,\,\leq\,\,19.3 \,\,\,&
\\[2.4mm]
\hline
\rule{0cm}{5mm}
\mbox{LO}
& \,\,\,8.0 \,\,\leq\,\,\varepsilon'/\varepsilon\,\,\leq\,\,62.1 \,\,\,&
\\[0.2mm]
\mbox{\,\,NDR\,\,}
& \,\,\,6.8 \,\,\leq\,\,\varepsilon'/\varepsilon\,\,\leq\,\,63.9 \,\,\,& 
\mbox{scanning}
\\[0.2mm]
\mbox{HV}
& \,\,\,2.8\,\,\leq\,\,\varepsilon'/\varepsilon\,\,\leq\,\,49.8 \,\,\, 
\\[2.4mm]
\hline  
\end{array}
\end{eqnarray*}
\end{table}

\section{ON THE SIZE OF THE ERRORS IN THE ANALYSIS of \boldmath 
$\varepsilon'/\varepsilon$ \unboldmath}

As shown in tabs.~\ref{tab1} and~\ref{tab2} the errors we obtain 
for $\varepsilon'/ \varepsilon$ are large. We believe however that the
uncertainties are not largely overestimated and reflect well our present 
knowledge on $\varepsilon' / \varepsilon$.

Note that we performed a scanning of the parameters. 
Refs.~\cite{bosch,buras2,BEF,CM2} used in addition to the scanning method
also a Monte Carlo procedure with gaussian distributions for the 
experimental input and flat distributions for the theoretical parameters. 
In this way a probability distribution is obtained for $\varepsilon'/ 
\varepsilon$, and the authors gave the median and the $68\,\%$ confidence 
level interval. We would like to emphasize that the use of this C.L.\
interval removes part of the hadronic uncertainties and leads to a range 
for $\varepsilon'/\varepsilon$ which is two times smaller than the one 
obtained from a full scanning over the ranges of the theoretical 
(mostly hadronic) parameters. In our opinion the Monte Carlo analysis
is misleading because there is no justification for assuming {\it any}
probability distribution for theoretical parameters, and it leads to an 
underestimate of the uncertainties in the calculation. To illustrate this 
point one might note that values for $\varepsilon'/\varepsilon$ above the 
$68\,\%$ C.L.\ range given e.g.\ in ref.~\cite{buras2} can be obtained for 
central values of the experimental parameters and for quite reasonable values 
of the theoretical ones within the ranges considered in this reference. Therefore
we think that a full scanning of the theoretical parameters, with gaussian 
distribution or full scanning for the experimental parameters\footnote{Since 
the ranges in tabs.~\ref{tab1} and~\ref{tab2} are predominantly due to the 
theoretical uncertainties, the difference between these two procedures is 
moderate.}, gives a better idea of the uncertainties in the CP ratio. A similar 
comment applies to the range for Im$\lambda_t$ since the determination of the CKM 
phase involves many non-perturbative theoretical parameters (as e.g.\ $\hat{B}_K$).

Moreover, as we explained above, non-chirally suppressed corrections 
beyond the large $N_c$ limit are essential for $B_6^{(1/2)}$. Therefore, 
in our opinion larger errors for $B_6^{(1/2)}$ should have been taken in 
ref.~\cite{buras2} since the authors did not calculate this parameter 
(in this case, from the counting in $p^2$ and $1/N_c$, the errors could be 
taken as large as $100\,\%$). The same comment applies to ref.~\cite{CM2} 
where a similar range for $\langle Q_6\rangle_0$ was adopted [$B_6^{(1/2)}$
was varied around its large-$N_c$ value taking an error of $100\,\%$, but 
$m_s$ was fixed adopting the value $(m_s + m_d)(2\,\mbox{GeV})=130\,
\mbox{MeV}$ without considering any error on it]. A similar statement
applies to the dispersive analysis of ref.~\cite{pich} which does 
not give access to any of the scale dependent non-factorizable terms 
(the non-chirally suppressed quadratic corrections in particular). 
The calculation of the scale dependent terms is not easier for lower 
values of the squared momentum of the kaon, and dispersive techniques
do not help in their calculation. To neglect these terms leads to a failure 
in reproducing the $\Delta I=1/2$ rule and is also not justified for 
$Q_6$.\footnote{Further comments on the dispersive analysis of the FSI 
effects in the calculation of $\varepsilon'/\varepsilon$ can be found in
ref.~\cite{abdis}.} 
As pointed out in ref.~\cite{CM2}, in the Chiral Quark Model \cite{BEF} 
the correlation between the $\Delta I=1/2$ amplitude and $\varepsilon'/
\varepsilon$ (used to fix the parameters necessary to estimate the 
matrix element of $Q_6$) is subject to potentially large uncertainties.
As for $\Omega_{\mbox{\tiny IB}}$, a minimum error of $\sim 0.10$ 
seems to be required for a careful estimate of $\varepsilon'/
\varepsilon$. The value $\Omega_{\mbox{\tiny IB}}=0.16\pm 0.03$ in 
ref.~\cite{ecker} (which has been used in ref.~\cite{CM2}) was obtained 
by investigating the $\pi-\eta$ contribution to this parameter (including 
$\eta-\eta'$ mixing). However, corrections beyond this term could be non-negligible 
as suggested by the numerical results of refs.~\cite{val,wolfe}.

We would like to emphasize also that the $\sim 25\,\%$ error obtained by 
comparing tabs.~\ref{tab1} and~\ref{tab2} should be included by any 
analysis which either does not include final state interactions or does 
not reproduce well the numerical values of the phases. This estimate of 
part of the neglected higher order corrections is usually not taken into 
account. 

We conclude that at present there is no method which can predict
$\varepsilon'/\varepsilon$ with an error much smaller than the one
presented in tabs.~\ref{tab1} and~\ref{tab2} which give a good idea 
of the uncertainties involved in the calculation of the CP ratio. 
The statement, that the experimental data can be accommodated (only) 
if all the hadronic parameters are taking values at the extreme of 
their reasonable ranges (see e.g.\ ref.~\cite{nir}), which is based on 
the use of the $68\,\%$ C.L.\ intervals of refs.~\cite{bosch,buras2,CM2}, 
has only weak theoretical foundations.

\section{COMMENTS ON \boldmath $\varepsilon'/\varepsilon$ \unboldmath}

Following our analysis, described above, a series of comments can be made:
\begin{itemize}
\item 
The use of the large-$N_c$ value or values close to it is not justified for 
$Q_6$ in $\varepsilon'/\varepsilon$ in the same way as for $Q_{1,2}$ in the 
$\Delta I = 1/2$ rule.
\item 
The related claim that we expect in general in the standard model a value 
of $\varepsilon'/\varepsilon$ smaller than the data is therefore not 
justified. One should note that the main methods used to calculate the 
$1/N_c$ corrections \cite{HKPS,BP2,BEF} all find them large and positive.
\item 
Dispersive techniques as proposed in ref.~\cite{pich} do not help in the 
calculation of many of these corrections (i.e.\ of the scale dependent terms).
\item 
Errors are large. A value between $\sim\mbox{few}\cdot 10^{-4}$ and 
$\sim 5\cdot 10^{-3}$ appears perfectly plausible without going beyond 
the reasonable ranges of the parameters.
\item 
Therefore $\varepsilon'/\varepsilon$ cannot be used at present to 
investigate new physics. Even if Im$\lambda_t$ was found as small 
as $0.6\cdot 10^{-4}$ as could be suggested by recent measurements 
\cite{babe}, we could still not exclude that the standard model 
reproduces the data.
\item 
At low energy there is a significant enhancement of $B_6^{(1/2)}$ 
\cite{HKPS}. It is quite reasonable that below $500\,-\,600\,\mbox{MeV}$ 
the model independent lagrangian of Eq.~(\ref{lag}) gives the bulk of the 
result. This is an important indication for a large value of $\varepsilon'/
\varepsilon$ in accordance with the data. The results of refs.~\cite{BP2,BEF} 
also point towards this direction.
\item 
The effects of dimension-eight operators could possibly change the results 
largely \cite{don2} and should certainly be investigated.
\item 
Despite the recent progress in the calculation of $\Omega_{\mbox{\tiny IB}}$
\cite{ecker,val,wolfe} the problem of an accurate calculation of this 
parameter is still relevant in the same way as for $B_6^{(1/2)}$.
\end{itemize}
{\bf Questions} (E.~Pallante, Univ.\ Barcelona):\\ 
1) {\it Can you comment on the one-loop estimate of $B_8^{(3/2)}$ 
in ref.~\cite{don}?} The results of ref.~\cite{don} (which have been obtained in 
the chiral limit) and our results for $B_8^{(3/2)}$ are not incompatible within 
their respective errors especially for moderate values of~$\Lambda_c$.\\
2) {\it The value of $\Omega_{\mbox{\tiny IB}}=0.16 \pm 0.03$ is 
actually not a pure large-$N_c$ calculation. There is no reason to expect 
large $1/N_c$ corrections in this case.} There is a priori no reason for
the $1/N_c$ corrections from irreducible one-loop diagrams beyond the 
reducible ones calculated in ref.~\cite{ecker} to be negligible (see 
also refs.~\cite{val,wolfe}).\\[3mm]
{\bf Acknowledgements}\\
We warmly thank our collaborators G.O.\ K\"ohler and E.A.\ Paschos. We
acknowledge partial support from BMBF, 057D093P(7), Bonn, FRG, DFG 
Antrag PA-10-1 and EEC, TMR-CT980169. 

\end{document}